\documentclass[final,3p,times,twocolumn]{elsarticle}

\usepackage[T1]{fontenc}
\usepackage[utf8]{inputenc}
\usepackage{lmodern}
\usepackage{hyperref}
\usepackage{lineno}
\usepackage{subfig}
\usepackage{graphicx}
\usepackage{threeparttable}
\usepackage{booktabs}
\usepackage[section]{placeins}

\usepackage{amsmath}
\usepackage[version=4]{mhchem}
\usepackage[table-align-text-post=false, range-units=single, range-phrase=\textendash, separate-uncertainty=false, multi-part-units=single, per-mode=symbol-or-fraction]{siunitx}

\journal{Journal of Molecular Physics}

\begin{document}
\begin{frontmatter}
\title{LLWP - A new Loomis-Wood Software at the Example of Acetone-\texorpdfstring{\ce{^{13}C1}}{13C1}}

\author[cologne]{Luis Bonah}
\author[cologne]{Oliver Zingsheim}
\author[cologne]{Holger S. P. Müller}
\author[rennes]{Jean-Claude Guillemin}
\author[cologne]{Frank Lewen}
\author[cologne]{Stephan Schlemmer}

\affiliation[cologne]{
    organization={I.\ Physikalisches Institut, Universität zu Köln},
    addressline={Zülpicher Str.\ 77}, 
    city={Köln},
    postcode={50937},
    country={Germany}
}
\affiliation[rennes]{
            organization={Univ Rennes, Ecole Nationale Supérieure de Chimie de Rennes},
            addressline={ISCR-UMR 6226}, 
            city={Rennes},
            postcode={35000}, 
            country={France}
}

\begin{abstract}
Acetone-\ce{^{13}C1} is a complex organic molecule with two internal methyl (\ce{-CH3}) rotors having relatively low effective barriers to internal rotation of about \SI{249}{cm^{-1}}.
This leads to two low-lying torsional modes and five internal rotation components resulting in a dense and complicated spectrum.
In this study, measurements of acetone-\ce{^{13}C1} were performed with an isotopically enriched sample in the frequency range \SIrange{37}{1102}{GHz}.
Predicted spectra of acetone-\ce{^{13}C1} created with ERHAM allow for future radio astronomical searches.

Loomis-Wood plots are one approach to improve and fasten the analysis of such crowded spectra.
In this study, the new Loomis-Wood software LLWP was used for fast and confident assignments.
LLWP focuses on being user-friendly, intuitive, and applicable to a broad range of assignment tasks.
The software is presented here and can be downloaded from \mbox{\texttt{\href{https://llwp.astro.uni-koeln.de}{llwp.astro.uni-koeln.de}}}.
\end{abstract}


\end{frontmatter}

\section{Introduction}
\label{sec:Introduction}
Acetone was first detected in Sgr B2(N) by Combes et al.~\cite{Combes1987} and later confirmed by Snyder et al.~\cite{Snyder2002}.
In the laboratory, first lines up to \SI{31}{\giga\hertz} were found by Bak et al.~\cite{Bak1949} and Weatherly and Williams~\cite{Weatherly1952}.
Swalen and Costain presented a structural study based on microwave spectroscopy~\cite{Swalen1959}.
Several subsequent studies investigated the spectrum of acetone, its isotopologs, and vibrationally excited states of the main isotopolog~\cite{Nelson1965, Peter1965, Vacherand1986, Groner2002, Groner2006, Groner2008, Lovas2006, Ilyushin2013, Ilyushin2019, Ordu2019}.

\begin{figure}[tb]
    \centering
    \includegraphics[width=1\linewidth]{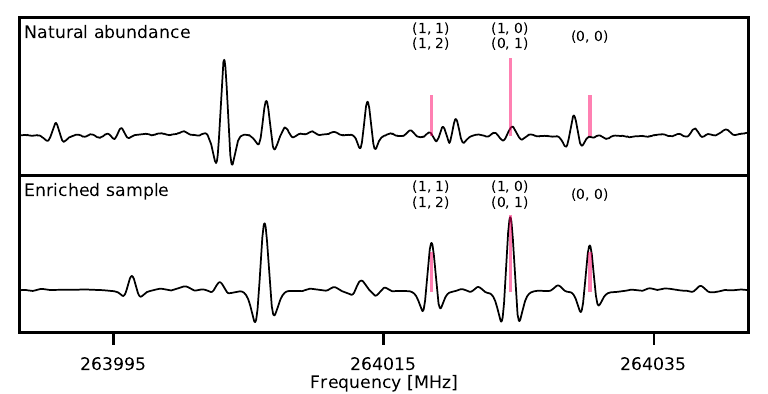}
    \caption{
		Spectrum and predictions (red sticks) of acetone-\ce{^{13}C1} at natural abundance (\SI{2.2}{\percent}, top) and for the enriched sample (\SI{98}{\percent}, bottom).
		The internal rotation components of the oblate paired transitions $27_{0,27} \leftarrow 26_{1,26}$ and $27_{1,27} \leftarrow 26_{0,26}$ are shown.
		For the enriched sample, the typical triplet pattern is clearly visible whereas at natural abundance all three peaks are blended and the pattern is not visible.
	}
    \label{fig:enriched}
\end{figure}

For the asymmetric \ce{^{13}C1} isotopolog (\ce{^{13}CH_3C(O)CH_3}), an initial analysis was presented by Lovas and Groner~\cite{Lovas2006} consisting of 55 transitions up to \SI{25}{\giga\hertz}.
Their results were extended by Ordu et al.~\cite{Ordu2019} to 110 transitions up to \SI{355}{\giga\hertz}.
The low number of assigned transitions in Ref.~\cite{Ordu2019}, compared to 9715 assigned transitions for an enriched sample of the symmetric \ce{CH_3 ^{13}C(O)CH_3} isotopolog in the same study, resulted from the lack of an enriched sample of the asymmetric species.
We revisited the spectrum with an enriched sample to circumvent difficulties tied to the low \SI{2.2}{\%} natural abundance of acetone-\ce{^{13}C1} (see \autoref{fig:enriched}).

Acetone-\ce{^{13}C1} has two inequivalent internal methyl (\ce{-CH3}) rotors with relatively low effective barriers to internal rotation.
A value of \SI{249.232}{cm^{-1}} was derived for the parent species\footnote{Variations for Acetone-\ce{^{13}C1} are assumed to be negligible.}~\cite{Ilyushin2019}.
In comparison to the main isotopolog, one methyl \ce{^{12}C} atom is substituted by a \ce{^{13}C} atom.
The coupling of the two distinguishable internal rotors with the overall rotation results in five internal rotation components labeled (following the nomenclature of Ref.~\cite{Lovas2006}) as $(0, 0)$, $(0, 1)$, $(1, 0)$, $(1, 1)$, and $(1, 2)$ which will be introduced in more detail later.

The internal rotation components and two energetically low-lying torsional modes lead to a dense and complicated spectrum.
Similar conditions can be found for many other complex molecules, with isotopologs, hyperfine structure, and other interactions being additional factors for a complex and line-rich spectrum.
Analyzing spectra close to the confusion limit, i.e., assigning lines unambiguously, proves to be cumbersome with conventional methods.
Approaches to accommodate this challenge include on the software side e.g., Loomis-Wood plots (LWPs)~\cite{Loomis1928}, Fortrat diagrams, and for infrared data the Automated Spectral Assignment Procedure~\cite{MartinDrumel2015b} as well as on the experimental side e.g., double-resonance spectroscopy~\cite{Stahl1984,Christen1995} and its advancement the double-modulation double-resonance spectroscopy~\cite{Zingsheim2021}.

LWPs display adjacent transitions of a series on top of each other. This makes it easy to follow series and identify deviations by using adjacent transitions as reference.
This results in more confident and efficient assignments.
Several programs using this approach exist in the literature, with Pgopher~\cite{Western2017}, the AABS package~\cite{Kisiel2005}, and LWW~\cite{Lodyga2007} being three popular options.
These programs are valuable tools, but unfortunately, they either lack different methods to determine center frequencies, use of experimental spectrum instead of peak lists, easy setup, support for multiple operating systems, documentation, or combinations of these.
In addition, the advancement in graphical user interface (GUI) libraries and computational power of personal computers alleviates some of the restrictions that previous generations of programs faced. 

Here LLWP is presented, a newly written Loomis-Wood plotting software for fast and confident assignment of real spectra.
It focuses on being intuitive and user-friendly while being proficient for a wide range of assignment tasks.
The analysis of the dense and complicated spectrum of acetone-\ce{^{13}C1} benefited immensely from the use of LLWP.

First, the experimental details are described (\autoref{sec:Experimental Details}), next the newly developed LLWP program is presented (\autoref{sec:The LLWP Program}), then the analysis of spectra of acetone-\ce{^{13}C1} is explained (\autoref{sec:Acetone Fit}), and finally the results are discussed and an outlook is given (\autoref{sec:Discussion}).

\section{Experimental Details}
\label{sec:Experimental Details}
Measurements were performed with a synthesized sample of acetone-\ce{^{13}C_1}.
High-resolution broadband spectra were measured in the frequency ranges \SIrange{37}{67}{\giga\hertz}, \SIrange{70}{129}{\giga\hertz}, and \SIrange{167}{1102}{\giga\hertz} using different experiments in Cologne, resulting in a total broadband coverage of \SI{1024}{\giga\hertz}.
First, the synthesis is described (\autoref{sec:Synthesis}) then the different experimental setups are presented (\autoref{sec:Experiments}).

\subsection{Synthesis of Acetone-\texorpdfstring{\ce{^{13}C_1}}{13C1}}
\label{sec:Synthesis}
The sample was synthesized using a procedure described by Winkel et al.~\cite{Winkel2010}.
In a \SI{50}{mL} round-bottomed flask equipped with a reflux condenser, a dropping funnel and a nitrogen inlet were introduced magnesium turnings (\SI{0.4}{g}, \SI{16.8}{mmol}, 1.2 equiv.) and \SI{15}{ml} of diethyl ether. \ce{^{13}C}-methyl iodide (\SI{2.0}{g}, \SI{14}{mmol}, 1 equiv.) diluted in \SI{10}{ml} of diethyl ether was added dropwise at such a rate that a gentle reflux was maintained.
The mixture was refluxed for \SI{10}{min} and then cooled at \SI{-10}{\celsius}.
Freshly distilled acetaldehyde (\SI{0.74}{g}; \SI{16.8}{mmol}; 1.2 equiv.) was slowly added, to give a white slurry and the mixture was stirred at room temperature for \SI{30}{min}.
The ether was evaporated and \SI{5}{ml} of water were carefully added.
Hydrochloric acid (6N) was added until pH 6.
The low boiling point compounds containing 2-propanol and water were distilled in vacuo (\SI{0.1}{mbar}) and then slowly added to a cold solution of sodium bichromate dihydrate (\SI{5.4}{g}, \SI{18}{mmol}, 1.3 equiv.) and sulfuric acid at 96\% (\SI{2.1}{g}) in \SI{3}{ml} of water.
The temperature should not exceed \SI{50}{\celsius}.
After \SI{10}{min} at room temperature, the acetone was purified by distillation in a vacuum line equipped with two U-traps.
The first immersed in a cold bath at \SI{-60}{\celsius} removed the high boiling point compounds and the second immersed in a liquid nitrogen bath selectively trapped the acetone.
This procedure yielded \SI{0.74}{g} (90\% yield) with 98\% enrichment of acetone-\ce{^{13}C1}.

\subsection{Experimental Setups}
\label{sec:Experiments}
Broadband measurements of the synthesized acetone-\ce{^{13}C1} sample were performed using three different experimental setups in Cologne~\cite{MartinDrumel2015, Ordu2019}.
Additionally, single measurements with longer integration times were performed in frequency ranges with low source output power\footnote{Transitions were ordered by the predicted intensity and the top 98 transitions in the frequency range \SIrange{750}{790}{\giga\hertz} as well as the top 1008 transitions in the frequency range \SIrange{37}{67}{\giga\hertz} were measured.}.
Especially the frequency range \SIrange{37}{67}{\giga\hertz} was important as in this low-frequency range the five internal rotation components are often resolved.
All experiments share a general structure, consisting of a source, an absorption cell, and a detector~\cite{MartinDrumel2015, Ordu2019}.
Horn antennas, lenses, and mirrors are used to guide the beam through the absorption cells and onto the detector.
The absorption cells are made out of borosilicate glass.
They are connected with a pump for evacuating the cell and with an inlet for the sample.
Different detector techniques, being Schottky detectors (<\SI{500}{GHz}) and a cryogenically cooled bolometer (>\SI{500}{GHz}), were used to optimize the SNR\@.
Lock-in amplifiers with a \textit{2f}-demodulation scheme were used.
As a result, lineshapes look similar to the second derivative of a Voigt profile.
All measurements were performed at room temperature and with gas pressures around \SI{10}{\micro bar}.

For \SIrange{37}{67}{\giga\hertz}, the signal is guided directly to the antenna and not multiplied, for \SIrange{70}{129}{\giga\hertz} a tripler is used.
For both frequency ranges, Schottky detectors are employed and connected to an in-house made bias box.
The absorption cell consists of two \SI{7}{\metre} glass cells in single-pass mode adding up to a total absorption path length of \SI{14}{\metre}.
The experimental setup is described in greater detail in Ref.~\cite{Ordu2019}.

The frequency range \SIrange{167}{515}{\giga\hertz} was covered with a commercially available source from Virginia Diodes Inc.\ (VDI) with three different setups consisting of cascaded doublers and triplers. As in the previous experiment, Schottky detectors were utilized.
The absorption cell of \SI{5}{\metre} length was used in double-pass mode resulting in a total absorption path length of \SI{10}{\metre}.
More detailed information can be found in Refs.~\cite{MartinDrumel2015, Ordu2019}.

The frequency range \SIrange{500}{1100}{\giga\hertz} was measured with two different setups of cascaded multipliers (VDI) and a QMC QNbB/PTC(2+XBI) hot-electron bolometer.
A single \SI{5}{\metre} cell was used in single-pass mode.

Standing waves were removed from the spectra by Fourier filtering using a self-written script\footnote{Download at \texttt{\url{https://github.com/Ltotheois/SnippetsForSpectroscopy/tree/main/FFTCorrection}}}.

\section{The LLWP Program}
\label{sec:The LLWP Program}
LLWP is presented, a newly developed software based on LWPs for exploring and assigning spectra.
It aims to be efficient, easy to handle, and in particular user-friendly. Additionally, LLWP offers great customizability and flexibility.
Distinguishing factors from previous programs~\cite{Western2017, Kisiel2005, Lodyga2007} are the use of real spectra instead of peak lists, and most importantly the simple setup and the improved user experience that is tied to the advancement in graphical user interface (GUI) libraries.
The software, its documentation, and contact information for feedback as well as feature requests can be found on LLWP's website\footnote{Visit \texttt{\url{https://llwp.astro.uni-koeln.de/}}}.

\begin{figure*}[tb]
    \centering
    \includegraphics[width=1\textwidth]{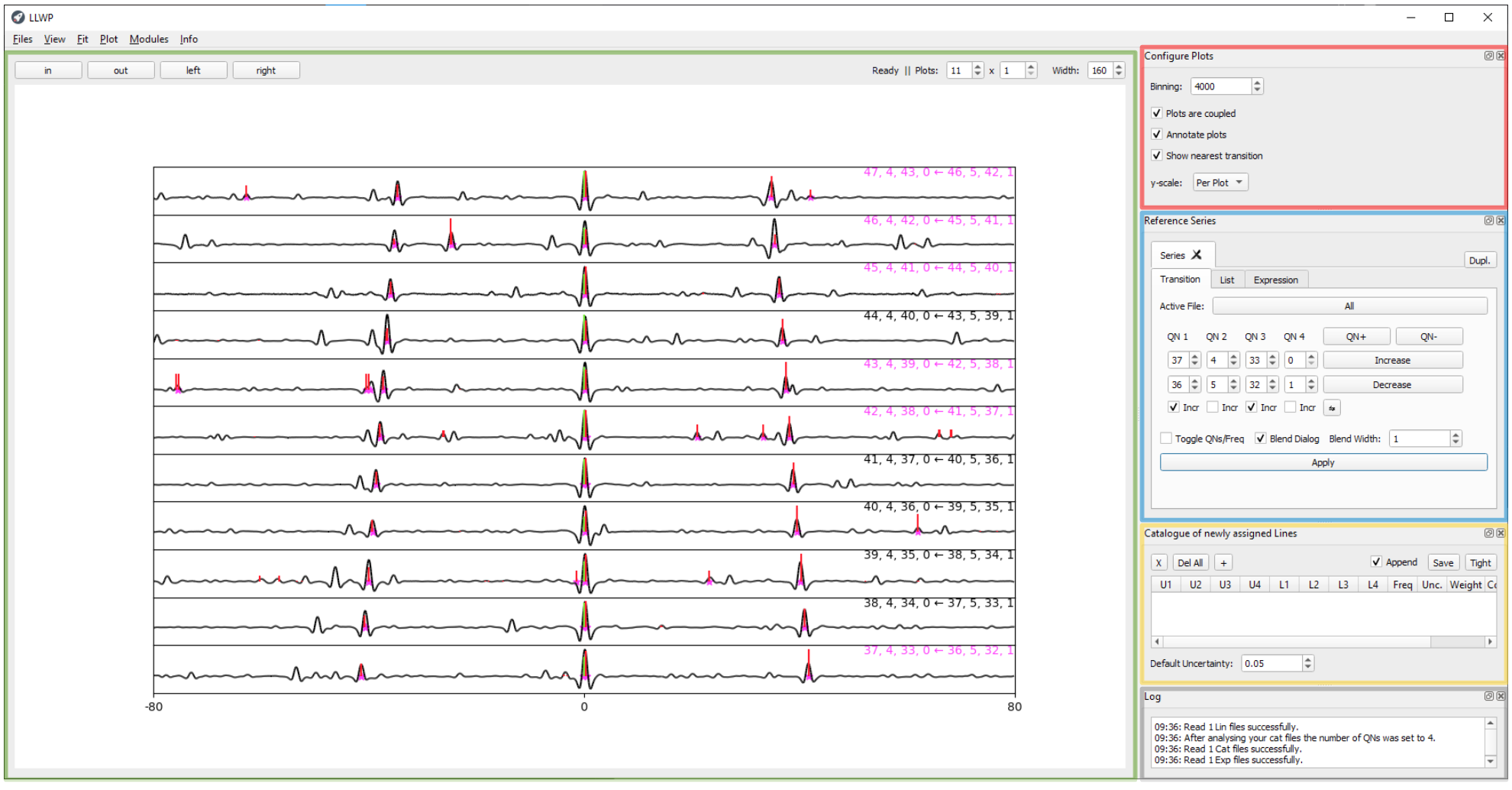}
    \caption{
		The GUI of the newly developed LLWP program.
		The five areas of the main window are highlighted by colored rectangles.
		The \textbf{Loomis-Wood plot} window (green rectangle) shows the experimental spectrum (black lines), the predictions (red sticks), and the reference series (green sticks).
		Already assigned transitions are indicated by pink stars and pink colored transitions (top right corner of the respective subplot).
		To the right of the LWP are the \textbf{Configure Plots} window (red rectangle) for options regarding the LWP, the \textbf{Reference Series} window (blue rectangle) which specifies the center positions of the LWP, the \textbf{Catalog of newly assigned Lines} (yellow rectangle) showing the assigned lines, and the \textbf{Log} window (gray rectangle) for messages, warnings, and errors.
		See the text for more information.
	}
    \label{fig:GUI LLWP}
\end{figure*}

First, the technical details are summarized (\autoref{sec:Technical Details}), then the core functionality is described in some detail to highlight the main advantages for potential users (\autoref{sec:Core Functionality}), and last a selection of additional features that support a thorough analysis are introduced to highlight the variety of tools provided by LLWP (\autoref{sec:Additional Features}).

\subsection{Technical Details}
\label{sec:Technical Details}
LLWP is written in python 3.
Therefore it runs on all operating systems that support python 3 and the required libraries.
For good performance, Pandas~\cite{McKinney2010} and NumPy~\cite{Harris2020} are used for data handling as they are written in C/C++.
The GUI uses PyQt5 and the fitting uses SciPy~\cite{Virtanen2020}.
Plots are created with Matplotlib~\cite{Hunter2007}.
For the handling of Pickett's \textit{*.cat}, \textit{*.lin}, and \textit{*.egy} formats~\cite{Pickett1991} our own Pyckett library was used\footnote{See \texttt{\url{https://pypi.org/project/pyckett/}} or install with pip via \texttt{pip install pyckett}}.
In addition to the python code, a Windows executable is provided on the website, making installation on Windows machines as simple as downloading a single file.

Certain tasks that are either computationally or I/O intensive are threaded to not block the user interface resulting in a responsive experience.
To keep performant with large datasets, each subplot plots only its currently visible data\footnote{The important step is to reduce the complete dataset to only the visible dataset before handing it to the plot function of the used library. For typical usecases, this drastically decreases computation time and memory consumption. However, for the user, apart from the mentioned performance benefits, there is no tangible difference between plotting all data but only showing an excerpt and only plotting the data in the excerpt}.
Additionally, even these subsets are downsampled if they exceed a user-selectable value.

\subsection{Core Functionality}
\label{sec:Core Functionality}
LLWP's core functionality is assigning center frequencies from a real spectrum to quantum numbers of predicted transitions.
LLWP relies on Loomis-Wood plots for increased confidence and speed.
The LWP consists of multiple subplots arranged in a vertical fashion.
Each subplot shows a region of the spectrum with the same width around its predicted center frequency.
The series of predicted center frequencies of all subplots is called reference series.
The easiest example would be a linear rotor which usually shows only small deviations from equidistantly spaced transitions with a distance of $2B$.
When using these ideal frequencies as the reference series, the LWP shows the deviations of the real transitions from this idealized model.
The trend in the LWP makes it easy to identify and follow the series and the row number corresponds to the $J$ quantum number of the upper state and makes assignment straight forward.
The software is not opinionated, meaning it is not limited to specific units or formats and works with \SIrange{1}{6}{} quantum numbers (or even no quantum numbers at all).
Additionally, LLWP can display multiple Loomis-Wood plots next to each other (see \autoref{fig:GUI LLWP COLS}).
Possible use cases are infrared data, e.g., examining $P$-, $Q$-, and $R$-branch series simultaneously, heavily perturbed systems, e.g., showing interaction partners side by side, or even as simple cases as comparing predictions from two different methods.
On the other extreme, LLWP can be used with only a single row and column to explore and assign spectra in a single plot, basically neglecting the Loomis-Wood character.
In summary, LLWP is a versatile tool for fast and confident assignments.

In the following, the most basic workflow is described, consisting of I) loading data, II) setting the reference series, III) assigning, and IV) saving the results\footnote{This basic workflow is presented in video format at \texttt{\url{https://llwp.astro.uni-koeln.de/videos/QuickGuide.mp4}}.}.
Simultaneously, the five areas of the main window (see \autoref{fig:GUI LLWP}) are introduced.

I) The first step is to load data into the program.
Three different types of data can be used in the program: i) files providing the experimental measurements, ii) files providing predictions, and iii) files providing already assigned lines.
All three data types can be added via the \textit{Files} menu or per drag-and-drop.
After this step, the \textbf{Loomis-Wood plot} (green rectangle in \autoref{fig:GUI LLWP}) shows a default series.
Important controls for the LWP are available on top, controlling the number of rows and columns as well as the width of the subplots.

II) In the second step, a user-defined reference series is selected in the \textbf{Reference Series} window (blue rectangle in \autoref{fig:GUI LLWP}).
Three options are available for specifying the center frequencies of the series, either by i) using catalog files and selecting a series, ii) via a list of frequencies, or iii) by entering a custom equation depending on $N$ (index of the subplot) and $N_0$ (offset index).
For the catalog files, Pickett's \textit{*.cat} format is supported by default but all fixed-width-formats (FWF) can be used.

III) Next, transitions of the reference series are assigned by selecting an area around the peak and the selected profile is fit to the selected data.
Currently available line profiles are a Gaussian, Lorentzian, and Voigt profile as well as their first and second derivatives.
Additionally, a polynomial with a selectable rank, a procedure testing polynomials with different ranks and using the best one, or a center-of-mass procedure adapted from Pgopher~\cite{Western2017} are available.
The determined center frequency and, if a catalog file is used, also the quantum numbers are added to the \textbf{Catalog of newly assigned Lines} window (yellow rectangle in \autoref{fig:GUI LLWP}).
Additionally, the uncertainty is set as i) a user-defined default value, ii) the absolute value of obs-calc, iii) user input to a dialog, or iv) the uncertainty from the fitting routine\footnote{Not all fitting routines support this option.
Use with caution.}.
In the \textbf{Catalog of newly assigned Lines} window, the assignments can be edited or deleted.
Already assigned lines are highlighted in the program to reduce confusion and prevent the user from unintentionally assigning the same line multiple times.
Steps II) and III) can be repeated for any desired reference series.

IV) When all targeted transitions are assigned, they are saved to a file, with Pickett's \textit{*.lin} format as the default format but all FWFs are available.

Additionally, the \textbf{Configure Plots} window (red rectangle in \autoref{fig:GUI LLWP}) provides further options for the appearance of the plot, with the most important being the scale of the LWP\@.
Either i) each plot is scaled individually, ii) all plots are scaled by the minimum and maximum of the whole spectrum, or iii) the user can set a custom scale.
For i) the predictions are also scaled individually whereas for ii) and iii) the predictions are scaled relative to the experimental spectrum.
Furthermore, the \textbf{Log} window (gray rectangle in \autoref{fig:GUI LLWP}) shows messages, warnings, and errors.

For convenience, all options can be set as default and the references to all currently opened files can be saved into a project eliminating the need of reloading all files individually.

\subsection{Additional Features}
\label{sec:Additional Features}
\begin{figure}[tb]
    \centering
    \includegraphics[width=1\linewidth]{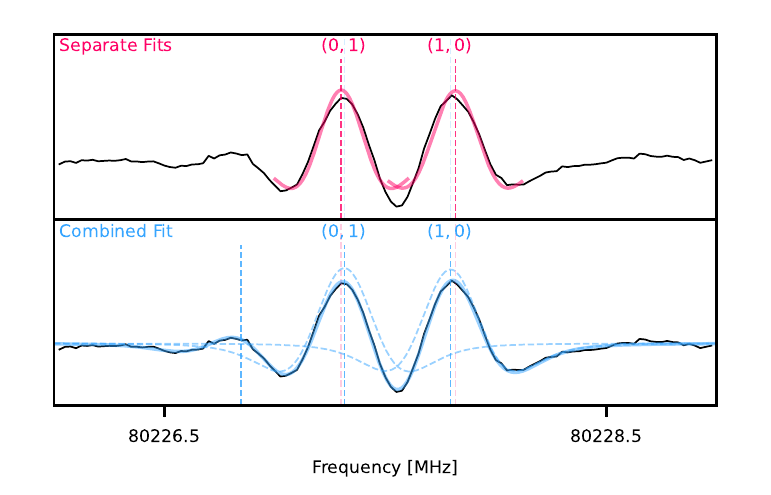}
    \caption{
		Example of the combined fit performed by the blended lines module (bottom in blue) in comparison with separate fits for each peak (top in red).
		The visible transitions are the $7_{2,6} \leftarrow	6_{1,5}$ transitions for the $(0,1)$ and $(1,0)$ internal rotation components.
		For the combined fit, an additional unknown peak at a lower frequency was included and the widths of the three peaks were locked.
		The center frequencies for each fit are indicated by dotted lines in the corresponding color and for the combined fit, the two components of interest are shown as dotted profiles.
		The center frequencies of the separate fit deviate by about \SI{20}{kHz}, emphasizing the advantage of a combined fit.
		For the analysis of acetone it was fundamental to accurately determine the splitting between $(0,1)$ and $(1,0)$ as well as between $(1,1)$ and $(1,2)$.
		However, even at low frequencies, these transitions overlap partly making a combined fit indispensable.
    }
    \label{fig:BlendedLinesModule}
\end{figure}

Several modules extend LLWP's core functionality.
Here, selected modules for an extended analysis, being the i) blended lines module, ii) seriesfinder module, iii) peakfinder module, and iv) residuals module, are presented with a short example of their respective use.
For more information on the modules visit the online documentation.

i) A major inconvenience for the assignment process are blended lines, as they make it complicated to determine precise center frequencies.
Many programs (e.g., ERHAM~\cite{Groner1997, Groner2012} and SPFIT~\cite{Pickett1991}) have built-in mechanisms to deal with blended lines.
For these cases, LLWP provides the ability to assign all blended lines to the center frequency of the blend and give each line its respective weight.
If this feature is activated and other predicted transitions are within a user-defined distance of the assigned transition, a dialogue window opens and the close-by transitions that should be assigned to the same blend can be selected.
On the other hand, partly blended lines can be resolved with the \textbf{blended lines module}.
It allows to fit multiple peaks simultaneously and thereby determine an individual center frequency for each transition.
The widths of the peaks can be independent or locked (e.g., if all peaks are from the same molecule and measurement and therefore are expected to have the same linewidth).
Additionally, a baseline can be subtracted in the form of a polynomial.
In the majority of cases, this results in more accurate center frequencies of partly blended transitions, see \autoref{fig:BlendedLinesModule}.

ii) The \textbf{seriesfinder module} allows to filter the predictions and display them ordered by intensity.
Predefined filters are present for the transition type, the frequency range, and to hide already assigned predictions.
One major use case is finding the strongest predicted, but so far unassigned transitions in the frequency range of the experiment.
For convenience, the transitions can be chosen as the reference series of the Loomis-Wood plot with a single click.

iii) The \textbf{peakfinder module} finds peaks in the real spectrum.
The peaks are shown in the spectrum and can be saved to a \textit{*.csv} file.
In addition, the peaks can be limited to so far unassigned peaks by providing a distance that peaks have to deviate from assigned transitions.
This allows to find the experimentally strongest, but yet unassigned peaks.

iv) Assigned transitions can be compared with predicted transitions in the \textbf{residuals module}.
By default, the residuals $\nu_{\text{Obs}}-\nu_{\text{Calc}}$ are shown against $\nu_{\text{Obs}}$ (see \autoref{fig: residuals}).
The residuals allow to assess the performance of the model and visually detect deviation patterns.
However, the $x$- and $y$-axis quantities can be chosen freely.
Therefore, a multitude of different plots can be created by the user, e.g., quantum number coverage plots (see \autoref{fig: qncoverage}) or weighted residuals (see \autoref{fig: weightedresiduals}).
For greater control, the transitions can be filtered and subgroups can be colored to highlight them in the plot.

To summarize, the modules complement the program's core capabilities and focus on more specific tasks.

\section{Spectroscopic Fingerprint of Acetone}
\label{sec:Acetone Fit}

Acetone-\ce{^{13}C_1} (\ce{^{13}CH_3C(O)CH_3}) is an asymmetric rotor with $\kappa = (2B-A-C)/(A-C) = 0.3150$, meaning acetone is an oblate rotor but far from the symmetric limit of +1.
The only nonzero dipole moment component of the main isotopolog is along the $b$-inertial axis with a value of \SI{2.93 \pm 0.03}{D}
\footnote{Acetone-\ce{^{13}C_1} is expected to have a small non-zero $a$-type dipole moment due to the \ce{^{13}C}-atom.
Thus, also $a$-type transitions are allowed.
However, the $a$-type dipole moment is negligibly small in comparison to the $b$-type dipole moment. Therefore, $a$-type transitions were neglected in the analysis.}~\cite{Peter1965}.
This results in a strong $b$-type spectrum with the selection rules $\Delta J = 0,\pm 1$ and $\Delta K_a= \pm 1, (\pm 3, ...)$ and $\Delta K_c= \pm 1, (\pm 3, ...)$.

\begin{figure}[tb]
    \centering
    \includegraphics[width=1\linewidth]{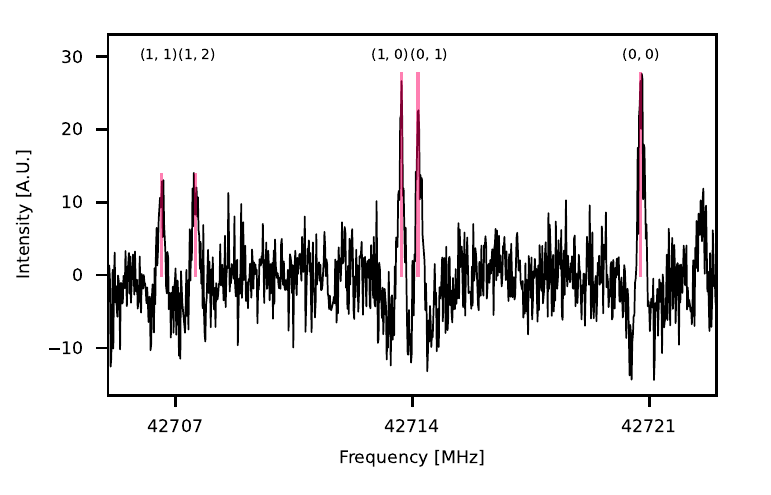}
    \caption{
		Acetone-\ce{^{13}C1} spectrum and predictions (red sticks) with resolved symmetry states.
		The $4_{0,4} \leftarrow 3_{1,3}$ transition is split into its five internal rotation components $(1, 1)$, $(1, 2)$, $(1, 0)$, $(0, 1)$, and $(0, 0)$ with an intensity ratio of 1:1:2:2:2.
    }
    \label{fig: resolvedsplitting}
\end{figure}

The two distinguishable internal methyl rotors lead to five internal rotation components which are labeled as $(\sigma_1, \sigma_2)$ with $\sigma_1$ and $\sigma_2$ being the symmetry numbers.
The nomenclature is adapted from Lovas and Groner~\cite{Lovas2006}, see their work for more information.
The five components are $(0, 0)$, $(0, 1)$, $(1, 0)$, $(1, 1)$, and $(1, 2)$.
For transitions with low frequencies, often all five internal rotation components are resolved with an intensity ratio of 1:1:2:2:2 (see \autoref{fig: resolvedsplitting}).
Single measurements of such resolved transitions were essential to accurately determine the energy tunneling parameters $\epsilon$ of the ERHAM model~\cite{Groner1997, Groner2012}.
For transitions with high quantum numbers, the internal rotation components $(0, 1)$ and $(1, 0)$ as well as $(1, 1)$ and $(1, 2)$ become degenerate, resulting in a typical triplet pattern with an intensity ratio of (1+1):(2+2):2 = 1:2:1 (see LWP of \autoref{fig:GUI LLWP}).

First assignments were straightforward due to the characteristic patterns and previous literature work~\cite{Lovas2006, Ordu2019}.
An iterative fitting procedure was used.
In each cycle, the model was updated and new improved predictions were calculated, which led to more assignments.
Assigning with LLWP was confident and efficient as multiple trends were visible in the LWPs.
The two most used trends were systematic deviation patterns ($\nu_\text{obs}-\nu_\text{calc}$) for adjacent rows and trends in typical patterns (see e.g., the typical triplet pattern becoming narrower with increasing frequency in \autoref{fig:GUI LLWP}).
These trends allowed to confidently assign transitions, even if they showed strong deviations from the initial predictions.

All literature transitions in the here measured frequency range were reassigned, especially as the transitions from Ordu et al.~\cite{Ordu2019} for the \ce{^{13}C1}-acetone isotopolog were misassigned above \SI{200}{GHz}.
For lower frequency areas that were not measured here, literature data were used consisting of 55 transitions in the frequency range \SIrange{10}{25}{GHz}~\cite{Lovas2006}.
Together with the here assigned lines, this resulted in a total of 16208 transitions with due to blends 8958 unique line frequencies, 
see \autoref{tab:transitiontypecoverage} for the transition type coverage.
Fits and predictions were performed with a modified version of ERHAM~\cite{Groner1997, Groner2012}.
Due to the size of the dataset, multiple array sizes were increased from the version available on the PROSPE webpage\footnote{Visit \texttt{\url{http://www.ifpan.edu.pl/~kisiel/prospe.htm}}}, e.g., the maximum of transitions was increased from 8191 to 16383, the number of tunneling parameters per state was increased from 37 to 199 and the number of predicted lines was increased from \SI{50000}{} to \SI{1000000}{}.

An inconvenience for the assignment process was label switching~\cite{Groner1998, Koerber2013, Endres2009}, which occurred especially for high $J$ values.
This behavior is described in ERHAM's manual and appears due to oblate paired transitions being degenerate.
A random phase is introduced to nonetheless diagonalize the matrix which randomly spreads the intensity between two allowed $b$-type transitions and two forbidden $a$-type transitions.
The $a$-type transitions were filtered out before fitting the assigned lines.
However, the randomly spread intensity can influence the center frequency of blends because LLWP automatically sets the weight according to the intensity from the \textit{*.cat} file.
This was corrected by manually setting the weights of affected transitions.
Another complication arose due to transitions affected by label switching changing between different fits.
Thus, certain transitions were not assigned in otherwise assigned series (see unassigned transitions in \autoref{fig:GUI LLWP}).
Additionally, specific lines were excluded from the fit due to either poor signal-to-noise ratio or being inseparably blended with unknown transitions.
However, the great majority of blends could be treated with LLWP's blended lines module (see \autoref{fig:BlendedLinesModule} and \autoref{sec:Additional Features}) or ERHAM's blend functionality.

Due to the high number of assignments, the uncertainties were assigned automatically.
A single second derivative Voigt profile was fit to the experimental spectrum for each assignment.
The mean deviation between the experimental and simulated lineshape was calculated via the root mean square (\textit{RMS}).
To normalize the deviation, this value was divided by the amplitude $A$ resulting in\footnote{A direct implementation in LLWP is planned for the future.}
\begin{equation}
    \textit{RMS}/A = \frac{1}{A}\sqrt{\frac{\sum_i (I_{\text{exp}, i} - I_{\text{fit}, i})^2}{N}}
\end{equation}
$I_{\text{exp}, i}$ and $I_{\text{fit},i}$ are the intensities at frequency $i$ of the experimental lineshape and fitfunction respectively, $N$ is the number of frequencies $i$ in the fit range, and $A$ the amplitude of the fitfunction.
This resulted in a quantity, which incorporates the SNR, the asymmetry of the line, and possible blends - all being aspects that are taken into account when assigning uncertainties manually.
The calculated $\textit{RMS}/A$ value was used to group the transitions into three classes with uncertainties of \SI{30}{\kilo\hertz}, \SI{50}{\kilo\hertz}, and \SI{70}{\kilo\hertz}.

Different approaches were tested for the fit.
First, a fit was created step-by-step with a script that added in every cycle the next best parameter to the fit.
However, a better result was obtained by adding whole sets of parameters at once and afterward removing parameters with high relative uncertainties.

\begin{table}[tb]
	\centering
	\caption{Molecular parameters for the ground state of \ce{^{13}CH_3C(O)CH_3} up to \SI{500}{\giga\hertz}}
	\label{tab: Parameters500GHz}

    \scalebox{0.57}{
	\begin{threeparttable}
\begin{tabular}{l l S[table-format=-5.8(2), round-mode = uncertainty] S[table-format=-5.8(2), round-mode = uncertainty]}
\toprule
                Parameter &             \space &                                \text{Rotor 1} &                                    \text{Rotor 2} \\
\midrule
                 $\rho_1$ &                    &         0.0599258678 \pm         0.0000096898 &         0.0613943095 \pm         0.0000195666 \\
                $\beta_1$ &      /\si{\degree} &        29.3671008010 \pm         0.0034946000 &        21.1450973950 \pm         0.0092657000 \\
                      $A$ &  /\si{\mega\hertz} & \multicolumn{2}{S[table-format=-5.8(2), round-mode = uncertainty]}{10083.0346220000 \pm         0.0001187031} \\
                      $B$ &  /\si{\mega\hertz} & \multicolumn{2}{S[table-format=-5.8(2), round-mode = uncertainty]}{ 8277.5073830000 \pm         0.0001045950} \\
                      $C$ &  /\si{\mega\hertz} & \multicolumn{2}{S[table-format=-5.8(2), round-mode = uncertainty]}{ 4811.4691220000 \pm         0.0000767709} \\
             $\Delta_{K}$ &  /\si{\kilo\hertz} & \multicolumn{2}{S[table-format=-5.8(2), round-mode = uncertainty]}{    9.4304387240 \pm         0.0007564589} \\
            $\Delta_{JK}$ &  /\si{\kilo\hertz} & \multicolumn{2}{S[table-format=-5.8(2), round-mode = uncertainty]}{   -2.7390328210 \pm         0.0004346925} \\
             $\Delta_{J}$ &  /\si{\kilo\hertz} & \multicolumn{2}{S[table-format=-5.8(2), round-mode = uncertainty]}{    4.5851382210 \pm         0.0001029727} \\
             $\delta_{K}$ &  /\si{\kilo\hertz} & \multicolumn{2}{S[table-format=-5.8(2), round-mode = uncertainty]}{   -0.3465922339 \pm         0.0002377324} \\
             $\delta_{J}$ &  /\si{\kilo\hertz} & \multicolumn{2}{S[table-format=-5.8(2), round-mode = uncertainty]}{    1.9153803080 \pm         0.0000456541} \\
               $\Phi_{K}$ &       /\si{\hertz} & \multicolumn{2}{S[table-format=-5.8(2), round-mode = uncertainty]}{    0.0323896205 \pm         0.0010091000} \\
              $\Phi_{KJ}$ &       /\si{\hertz} & \multicolumn{2}{S[table-format=-5.8(2), round-mode = uncertainty]}{    0.0315216055 \pm         0.0009954621} \\
              $\Phi_{JK}$ &       /\si{\hertz} & \multicolumn{2}{S[table-format=-5.8(2), round-mode = uncertainty]}{   -0.0273226050 \pm         0.0003628382} \\
               $\Phi_{J}$ &       /\si{\hertz} & \multicolumn{2}{S[table-format=-5.8(2), round-mode = uncertainty]}{    0.0051273379 \pm         0.0000509359} \\
               $\phi_{K}$ &       /\si{\hertz} & \multicolumn{2}{S[table-format=-5.8(2), round-mode = uncertainty]}{   -0.0750360789 \pm         0.0003536566} \\
              $\phi_{JK}$ &       /\si{\hertz} & \multicolumn{2}{S[table-format=-5.8(2), round-mode = uncertainty]}{    0.0354140392 \pm         0.0001874593} \\
               $\phi_{J}$ &       /\si{\hertz} & \multicolumn{2}{S[table-format=-5.8(2), round-mode = uncertainty]}{    0.0025961052 \pm         0.0000245604} \\
       $\epsilon_{ 1 1 }$ &  /\si{\mega\hertz} & \multicolumn{2}{S[table-format=-5.8(2), round-mode = uncertainty]}{    1.0305751840 \pm         0.0075314000} \\
      $\epsilon_{ 1 -1 }$ &  /\si{\mega\hertz} & \multicolumn{2}{S[table-format=-5.8(2), round-mode = uncertainty]}{    0.0944118741 \pm         0.0012839000} \\
       $\epsilon_{ 1 0 }$ &  /\si{\mega\hertz} &        -758.0194611 \pm         0.0355453 &        -763.2298877 \pm         0.0317754 \\
       $\epsilon_{ 2 0 }$ &  /\si{\mega\hertz} &         0.7627163725 \pm         0.0026362000 &         0.7720957999 \pm         0.0023474000 \\
    $[A-(B+C)/2]_{ 1 0 }$ &  /\si{\mega\hertz} &         0.0319716785 \pm         0.0004939046 &         0.0371082436 \pm         0.0006157476 \\
      $[(B+C)/2]_{ 1 0 }$ &  /\si{\kilo\hertz} &         3.3733914510 \pm         0.2547436000 &        -3.6773779170 \pm         0.4282350000 \\
      $[(B-C)/4]_{ 1 0 }$ &  /\si{\kilo\hertz} &         9.0148627830 \pm         0.1282679000 &         5.1164067410 \pm         0.2159953000 \\
 $-[\Delta_{ K }]_{ 1 0 }$ &  /\si{\kilo\hertz} &         0.0243135840 \pm         0.0003563736 &         0.0134861001 \pm         0.0003380171 \\
$-[\Delta_{ JK }]_{ 1 0 }$ &  /\si{\kilo\hertz} &        -0.0257639433 \pm         0.0003655198 &        -0.0158379871 \pm         0.0003593501 \\
 $-[\Delta_{ J }]_{ 1 0 }$ &       /\si{\hertz} &         1.6618377530 \pm         0.1229201000 &         2.2402824410 \pm         0.1294602000 \\
$-[\delta_{ KJ }]_{ 1 0 }$ &  /\si{\kilo\hertz} &        -0.0181971379 \pm         0.0001137156 &        -0.0117632448 \pm         0.0001023934 \\
$-[\delta_{ JK }]_{ 1 0 }$ &       /\si{\hertz} &         0.7816331100 \pm         0.0621238000 &         1.2716915770 \pm         0.0654573000 \\
$[B^-_{ 0 0 1 }]_{ 1 0 }$ &  /\si{\mega\hertz} &        -0.3297735401 \pm         0.0053658000 &        -0.3140347530 \pm         0.0087563000 \\
$[B^-_{ 0 1 0 }]_{ 1 0 }$ &  /\si{\mega\hertz} &        -0.6409901970 \pm         0.0140296000 &         0.8611737639 \pm         0.0295473000 \\
$[B^-_{ 0 1 2 }]_{ 1 0 }$ &  /\si{\kilo\hertz} &         0.1783855073 \pm         0.0047549000 &        -0.0682767422 \pm         0.0066396000 \\
$[B^-_{ 0 3 0 }]_{ 1 0 }$ &  /\si{\kilo\hertz} &         0.3080397752 \pm         0.0098622000 &        -0.7231139971 \pm         0.0117915000 \\
$[B^-_{ 2 1 0 }]_{ 1 0 }$ &  /\si{\kilo\hertz} &        -0.5487087145 \pm         0.0078515000 &         0.7397115755 \pm         0.0120159000 \\
\midrule
Transitions &                   & \multicolumn{2}{S[table-format=-5.8]}{8125}  \\
Lines       &                   & \multicolumn{2}{S[table-format=-5.8]}{5819}  \\
\textit{RMS}    &    /\si{\mega\hertz}               & \multicolumn{2}{S[table-format=-5.8]}{0.053938}  \\ 
\textit{WRMS}        &                   & \multicolumn{2}{S[table-format=-5.8]}{1.136331} \\ 
\bottomrule
\end{tabular}
\begin{tablenotes}
\item \textbf{{Note.}} Fits performed with ERHAM in A-reduction.
Standard errors are given in parentheses.
Parameters are given in notation for the first rotor.
Interaction parameters are given in $B^-_{kpr}$ notation.
The minus sign indicates, that $\omega = -1$, while $k$, $p$, and $r$ are the powers of the operators $\text{P}^k$, $\text{P}_z^p$, and $\text{P}_+^r+\text{P}_-^r$, respectively.
\end{tablenotes}
\end{threeparttable}}
\end{table}

Up to \SI{500}{\giga\hertz}, fitting was straightforward.
The previous step-by-step fit up to \SI{500}{\giga\hertz} was used to get an approximation for which orders of parameters were needed.
Except for fundamental parameters\footnote{For both fits $\alpha_1$ and $\alpha_2$ were fixed to zero while $\rho_1$, $\rho_2$, $\beta_1$, $\beta_2$, $A$, $B$, $C$, $\epsilon_{01}$ and $\epsilon_{10}$ had sensible initial values from previous fits.}, all other parameters were given an initial value of zero.
Different sets of interaction parameters were tested.
The resulting fit includes 8125 transitions (5819 unique lines) with the covered quantum numbers being shown in \autoref{fig: qncoverage500GHz}.
The \textit{WRMS} is \SI{1.14}{} and the highest parameter uncertainty is about 11\%, so all parameters were kept.
The resulting parameters are shown in \autoref{tab: Parameters500GHz} and the residuals of the final fit are shown in \autoref{fig: residuals500GHz}.

\begin{table}[tb]
	\centering
	\caption{Molecular parameters for the ground state of \ce{^{13}CH_3C(O)CH_3} up to \SI{1100}{\giga\hertz}}
	\label{tab: Parameters1200GHz}
    \scalebox{0.57}{
	\begin{threeparttable}
\begin{tabular}{l l S[table-format=-5.8(2), round-mode = uncertainty] S[table-format=-5.8(2), round-mode = uncertainty]}
\toprule
                Parameter &             \space &                                \text{Rotor 1} &                              \text{Rotor 2}\\
\midrule
                   $\rho$ &                    &         0.0595398513 \pm         0.0000039301 &         0.0610286067 \pm         0.0000083300 \\
                  $\beta$ &      /\si{\degree} &        29.4232190560 \pm         0.0018207000 &        21.2919705910 \pm         0.0062719000 \\
                      $A$ &  /\si{\mega\hertz} & \multicolumn{2}{S[table-format=-5.8(2), round-mode = uncertainty]}{ 10083.0334610000 \pm         0.0000795969 } \\
                      $B$ &  /\si{\mega\hertz} & \multicolumn{2}{S[table-format=-5.8(2), round-mode = uncertainty]}{  8277.5065070000 \pm         0.0000962534 } \\
                      $C$ &  /\si{\mega\hertz} & \multicolumn{2}{S[table-format=-5.8(2), round-mode = uncertainty]}{  4811.4687230000 \pm         0.0000580234 } \\
             $\Delta_{K}$ &  /\si{\kilo\hertz} & \multicolumn{2}{S[table-format=-5.8(2), round-mode = uncertainty]}{     9.4281645510 \pm         0.0002358291 } \\
            $\Delta_{JK}$ &  /\si{\kilo\hertz} & \multicolumn{2}{S[table-format=-5.8(2), round-mode = uncertainty]}{    -2.7398614730 \pm         0.0002537800 } \\
             $\Delta_{J}$ &  /\si{\kilo\hertz} & \multicolumn{2}{S[table-format=-5.8(2), round-mode = uncertainty]}{     4.5853164290 \pm         0.0000650477 } \\
             $\delta_{K}$ &  /\si{\kilo\hertz} & \multicolumn{2}{S[table-format=-5.8(2), round-mode = uncertainty]}{    -0.3431673536 \pm         0.0001314881 } \\
             $\delta_{J}$ &  /\si{\kilo\hertz} & \multicolumn{2}{S[table-format=-5.8(2), round-mode = uncertainty]}{     1.9153245070 \pm         0.0000322407 } \\
               $\Phi_{K}$ &       /\si{\hertz} & \multicolumn{2}{S[table-format=-5.8(2), round-mode = uncertainty]}{     0.0288296584 \pm         0.0001853648 } \\
              $\Phi_{KJ}$ &       /\si{\hertz} & \multicolumn{2}{S[table-format=-5.8(2), round-mode = uncertainty]}{     0.0314096756 \pm         0.0002650126 } \\
              $\Phi_{JK}$ &       /\si{\hertz} & \multicolumn{2}{S[table-format=-5.8(2), round-mode = uncertainty]}{    -0.0269546094 \pm         0.0001371650 } \\
               $\Phi_{J}$ & /\si{\milli\hertz} & \multicolumn{2}{S[table-format=-5.8(2), round-mode = uncertainty]}{     5.5307991410 \pm         0.0256480000 } \\
               $\phi_{K}$ &       /\si{\hertz} & \multicolumn{2}{S[table-format=-5.8(2), round-mode = uncertainty]}{    -0.0684281272 \pm         0.0001079546 } \\
              $\phi_{JK}$ &       /\si{\hertz} & \multicolumn{2}{S[table-format=-5.8(2), round-mode = uncertainty]}{     0.0356155101 \pm         0.0000498528 } \\
               $\phi_{J}$ & /\si{\milli\hertz} & \multicolumn{2}{S[table-format=-5.8(2), round-mode = uncertainty]}{     2.7486668500 \pm         0.0128294000 } \\
     $[l_{ JK }]_{ 0 0 }$ & /\si{\micro\hertz} & \multicolumn{2}{S[table-format=-5.8(2), round-mode = uncertainty]}{     0.3461113359 \pm         0.0047042000 } \\
       $\epsilon_{ 1 1 }$ &  /\si{\mega\hertz} & \multicolumn{2}{S[table-format=-5.8(2), round-mode = uncertainty]}{     1.0890665880 \pm         0.0091778000 } \\
      $\epsilon_{ 1 -1 }$ &  /\si{\mega\hertz} & \multicolumn{2}{S[table-format=-5.8(2), round-mode = uncertainty]}{     0.1063782795 \pm         0.0015670000 } \\
       $\epsilon_{ 1 0 }$ &  /\si{\mega\hertz} &      -758.1742287120 \pm         0.0257209000 &      -763.2530793970 \pm         0.0217815000 \\
       $\epsilon_{ 2 0 }$ &  /\si{\mega\hertz} &         0.7862081700 \pm         0.0029648000 &         0.7568203116 \pm         0.0026515000 \\

    $[A-(B+C)/2]_{ 1 0 }$ &  /\si{\mega\hertz} &         0.0189214914 \pm         0.0002684708 &         0.0324093735 \pm         0.0003642869 \\
      $[(B+C)/2]_{ 1 0 }$ &  /\si{\mega\hertz} &         0.0102912501 \pm         0.0001337436 &        -0.0036745513 \pm         0.0002973803 \\
      $[(B-C)/4]_{ 1 0 }$ &  /\si{\mega\hertz} &         0.0122969635 \pm         0.0000691222 &         0.0049140212 \pm         0.0001519602 \\
 $[\Delta_{ K }]_{ 1 0 }$ &  /\si{\kilo\hertz} &         0.0176619284 \pm         0.0002090740 &         0.0111897270 \pm         0.0002065718 \\
$[\Delta_{ JK }]_{ 1 0 }$ &  /\si{\kilo\hertz} &        -0.0170935762 \pm         0.0001726945 &        -0.0129334334 \pm         0.0002112808 \\
 $[\Delta_{ J }]_{ 1 0 }$ &       /\si{\hertz} &        -2.0050137070 \pm         0.0592397000 &         1.9643869720 \pm         0.0841131000 \\
$[\delta_{ KJ }]_{ 1 0 }$ &  /\si{\kilo\hertz} &        -0.0166685944 \pm         0.0000556668 &        -0.0100328461 \pm         0.0000567814 \\
$[\delta_{ JK }]_{ 1 0 }$ &       /\si{\hertz} &        -0.8241532544 \pm         0.0301345000 &         1.1527885540 \pm         0.0427100000 \\
   $[\Phi_{ K }]_{ 1 0 }$ & /\si{\milli\hertz} &         5.8769578410 \pm         0.0656941000 &         1.3571572300 \pm         0.0585412000 \\
  $[\Phi_{ KJ }]_{ 1 0 }$ &       /\si{\hertz} &        -0.0111083112 \pm         0.0000672610 &        -0.0026788542 \pm         0.0000576606 \\
  $[\Phi_{ JK }]_{ 1 0 }$ & /\si{\milli\hertz} &         5.1390342570 \pm         0.0285161000 &         1.1431751800 \pm         0.0264631000 \\
  $[\phi_{ JK }]_{ 1 0 }$ & /\si{\milli\hertz} &         2.0316377380 \pm         0.0181494000 &         0.0975351343 \pm         0.0167926000 \\
$[B^-_{ 0 0 1 }]_{ 1 0 }$ &  /\si{\mega\hertz} &        -0.4124810937 \pm         0.0058749000 &        -0.3086527072 \pm         0.0076439000 \\
$[B^-_{ 0 1 0 }]_{ 1 0 }$ &  /\si{\mega\hertz} &        -1.1989098560 \pm         0.0054732000 &         1.4980968980 \pm         0.0102272000 \\
$[B^-_{ 0 1 2 }]_{ 1 0 }$ &  /\si{\kilo\hertz} &         0.3799690050 \pm         0.0015119000 &        -0.2078679530 \pm         0.0020430000 \\
$[B^-_{ 0 3 0 }]_{ 1 0 }$ &  /\si{\kilo\hertz} &         0.0374618283 \pm         0.0062587000 &        -0.4357750561 \pm         0.0060080000 \\
$[B^-_{ 2 1 0 }]_{ 1 0 }$ &  /\si{\kilo\hertz} &        -0.2595135992 \pm         0.0028113000 &         0.3991460781 \pm         0.0037218000 \\
$[B^-_{ 0 5 0 }]_{ 1 0 }$ &       /\si{\hertz} &        -0.2818237953 \pm         0.0024230000 &         0.0493392525 \pm         0.0021690000 \\
$[B^-_{ 2 3 0 }]_{ 1 0 }$ &       /\si{\hertz} &         0.3100482675 \pm         0.0014802000 &        -0.0632094686 \pm         0.0014602000 \\
$[B^-_{ 4 1 0 }]_{ 1 0 }$ &       /\si{\hertz} &        -0.0299206235 \pm         0.0004366786 &         0.0272320853 \pm         0.0004476520 \\
\midrule
Transitions     &                   & \multicolumn{2}{S[table-format=-5.8]}{12403}  \\
Lines     &                   & \multicolumn{2}{S[table-format=-5.8]}{8602}  \\
\textit{RMS}  &       /\si{\mega\hertz}           & \multicolumn{2}{S[table-format=-5.8]}{0.068270}  \\ 
\textit{WRMS}      &                   & \multicolumn{2}{S[table-format=-5.8]}{1.452576} \\ 
\bottomrule
\end{tabular}
\begin{tablenotes}
\item \textbf{{Note.}} Fits performed with ERHAM in A-reduction.
Standard errors are given in parentheses.
Parameters are given in notation for the first rotor.
Interaction parameters are given in $B^-_{kpr}$ notation.
The minus sign indicates, that $\omega = -1$, while $k$, $p$, and $r$ are the powers of the operators $\text{P}^k$, $\text{P}_z^p$, and $\text{P}_+^r+\text{P}_-^r$ respectively.
\end{tablenotes}
\end{threeparttable}}
\end{table}

For frequencies higher than \SI{500}{\giga\hertz}, LLWP's residuals module was used to identify a group of transitions with high quantum numbers that showed strong deviations.
The affected transitions with $94 > J' > 67$ and $K_a' > 3$ were excluded from the final fit.
Their deviation patterns look similar to patterns caused by interactions but were not further examined here (see \autoref{fig: residuals1200GHz}).
We expect more transitions with high quantum numbers to be perturbed and thus the fit to be effective. Consequentially, it should not be used for structure determination.
The resulting fit up to \SI{1100}{\giga\hertz} includes 12403 transitions (8602 unique lines) with a quantum number coverage as shown in \autoref{fig: qncoverage1200GHz}.
As with the previous fit, sets of parameters were added together, resulting in a fit with many parameters having high uncertainties, some even greater than 100\%.
Parameters were then omitted symmetrically for the two rotors.
Their uncertainty and influence on the goodness of the fit were the two criteria for choosing the next parameter to omit.
This allowed to reduce the number of parameters to 66 and improve the relative uncertainties.
For the final fit, the highest relative uncertainty is 17\% and only 2 parameters have uncertainties higher than 10\%.
The resulting \textit{WRMS} is \SI{1.45}{}.
The resulting parameters are shown in \autoref{tab: Parameters1200GHz} and the residuals of the final fit are shown in \autoref{fig: residuals1200GHz}.

\begin{table*}[tb]
	\centering
	\caption{Selected spectroscopic parameters for the fits of \ce{CH3C(O)CH3}, \ce{CH3 ^{13}C(O)CH3}, and \ce{^{13}CH3C(O)CH3} from Ordu et al.~\cite{Ordu2019} as well as the two fits for \ce{^{13}CH3C(O)CH3} from this work.}
	\label{tab: ParameterComparison}

	\scalebox{0.7}{
	\begin{threeparttable}
	\begin{tabular}{l l *{6}{S[table-format=-5.8(2), round-mode = uncertainty]}}
		\toprule
		\multicolumn{2}{l}{} & \multicolumn{3}{c}{Ordu et al.~\cite{Ordu2019}} & \hspace{0.35cm} & \multicolumn{2}{c}{This work} \\
		\multicolumn{2}{l}{Parameter}          &  \text{\ce{CH3C(O)CH3}} &  \text{\ce{CH3 ^{13}C(O)CH3}} &  \text{\ce{^{13}CH3C(O)CH3}} & &  \text{\ce{^{13}CH3C(O)CH3}} &  \text{\ce{^{13}CH3C(O)CH3}} \\
		\midrule                            
		$\rho_1$        &                       &   0.0619535(120)        &   0.0614858(41)               &   0.060458(57)               & &   0.0599259(97)              &   0.0595399(39)              \\
		$\rho_2$        &                       &                         &                               &   0.061811(58)               & &   0.061394(20)               &   0.0610286(83)              \\
		$\beta_1$       & /\si{\degree}         &   25.5065(76)           &   25.7140(31)                 &   29.5967(77)                & &   29.3671(35)                &   29.4232(18)                \\
		$\beta_2$       & /\si{\degree}         &                         &                               &   21.2614(76)                & &   21.1451(93)                &   21.2920(63)                \\
                    $A$ & /\si{\mega\hertz} &   10165.217780(280)     &   10164.005782(145)           &   10083.03218(58)            & &   10083.03462(12)            &   10083.033461(80)           \\
                    $B$ & /\si{\mega\hertz} &   8515.163068(248)      &   8516.083092(104)            &   8277.50617(44)             & &   8277.50738(10)             &   8277.506 507(96)           \\
                    $C$ & /\si{\mega\hertz} &   4910.198777(209)      &   4910.235399(107)            &   4811.468965(174)           & &   4811.469122(77)            &   4811.468 723(58)           \\
           $\Delta_{K}$ & /\si{\kilo\hertz} &   9.79105(88)           &   9.85055(33)                 &   9.456(60)                  & &   9.43044(76)                &   9.42816(24)                \\
          $\Delta_{JK}$ & /\si{\kilo\hertz} &   -3.17067(60)          &   -3.192449(194)              &   -2.865(74)                 & &   -2.73903(43)               &   -2.73986(25)               \\
           $\Delta_{J}$ & /\si{\kilo\hertz} &   4.854449(219)         &   4.852046(73)                &   4.5557(147)                & &   4.58514(10)                &   4.585316(65)               \\
           $\delta_{K}$ & /\si{\kilo\hertz} &   -0.60720(39)          &   -0.619619(127)              &   -0.226(31)                 & &   -0.34659(24)               &   -0.34317(13)               \\
           $\delta_{J}$ & /\si{\kilo\hertz} &   2.038605(99)          &   2.0381002(265)              &   1.8855(77)                 & &   1.915380(46)               &   1.915325(32)               \\
     $\epsilon_{ 1 0 }$ & /\si{\mega\hertz} &   -764.737(38)          &   -763.9260(53)               &   -759.25(81)                & &   -758.019(36)               &   -758.174(26)               \\
     $\epsilon_{ 0 1 }$ & /\si{\mega\hertz} &                         &                               &   -766.18(79)                & &   -763.230(32)               &   -763.253(22)               \\
     $\epsilon_{ 2 0 }$ & /\si{\mega\hertz} &   0.77490(283)          &   0.76612(173)                &   0.574(45)                  & &   0.7627(26)                 &   0.7862(30)                  \\
     $\epsilon_{ 0 2 }$ & /\si{\mega\hertz} &                         &                               &   0.574(45) \tnote{a}        & &   0.7721(23)                 &   0.7568(27)                  \\
     $\epsilon_{ 1 1 }$ & /\si{\mega\hertz} &   1.0902(95)            &   1.1059(61)                  &   1.0000(212)                & &   1.0306(75)                 &   1.0891(92)                 \\
     $\epsilon_{ 1-1 }$ & /\si{\mega\hertz} &   0.07346(262)          &   0.08853(176)                &   0.08630(143)               & &   0.0944(13)                 &   0.1064(16)                  \\
		\midrule
		\multicolumn{2}{l}{Transitions}                     &   \multicolumn{1}{S[table-format=-5.2]}{2181}                  &   \multicolumn{1}{S[table-format=-5.2]}{9715    }                    &   \multicolumn{1}{S[table-format=-5.2]}{110     }                  &  & \multicolumn{1}{S[table-format=-5.2]}{8125    }                  &   \multicolumn{1}{S[table-format=-5.2]}{12403   }                  \\
		\multicolumn{2}{l}{Lines}                           &   \multicolumn{1}{S[table-format=-5.2]}{1862}                  &   \multicolumn{1}{S[table-format=-5.2]}{5870    }                    &   \multicolumn{1}{S[table-format=-5.2]}{72      }                  &  & \multicolumn{1}{S[table-format=-5.2]}{5819    }                  &   \multicolumn{1}{S[table-format=-5.2]}{8602    }                  \\
		\multicolumn{2}{l}{\textit{RMS} /\si{\kilo\hertz}}  &   \multicolumn{1}{S[table-format=-5.2, round-mode=places, round-precision=0]}{103.999}              &   \multicolumn{1}{S[table-format=-5.2, round-mode=places, round-precision=0]}{110.596}                    &   \multicolumn{1}{S[table-format=-5.2, round-mode=places, round-precision=0]}{29.381}                  & &  \multicolumn{1}{S[table-format=-5.2, round-mode=places, round-precision=0]}{53.938}                  &   \multicolumn{1}{S[table-format=-5.2, round-mode=places, round-precision=0]}{68.270}                  \\
		\multicolumn{2}{l}{\textit{WRMS}}                   &   \multicolumn{1}{S[table-format=-5.2, round-mode=places, round-precision=2]}{0.930755}              &   \multicolumn{1}{S[table-format=-5.2, round-mode=places, round-precision=2]}{1.280520}                    &   \multicolumn{1}{S[table-format=-5.2, round-mode=places, round-precision=2]}{0.748306}                  &  & \multicolumn{1}{S[table-format=-5.2, round-mode=places, round-precision=2]}{1.136331}                  &   \multicolumn{1}{S[table-format=-5.2, round-mode=places, round-precision=2]}{1.452576}                  \\
		\multicolumn{2}{l}{Standard Deviation\tnote{b}}       &   \multicolumn{1}{S[table-format=-5.2, round-mode=places, round-precision=2]}{0.95}                  &   \multicolumn{1}{S[table-format=-5.2, round-mode=places, round-precision=2]}{1.29    }                    &   \multicolumn{1}{S[table-format=-5.2, round-mode=places, round-precision=2]}{0.93    }                  & &  \multicolumn{1}{S[table-format=-5.2, round-mode=places, round-precision=2]}{1.11    }                  &   \multicolumn{1}{S[table-format=-5.2, round-mode=places, round-precision=2]}{1.46    }                  \\
		\bottomrule
	\end{tabular}
	\begin{tablenotes}\footnotesize
	\item \textbf{{Note.}} Fits performed with ERHAM in A-reduction.
Standard errors are given in parentheses.
	\item[a] Parameter of $(q, q') = (0, 2)$ is fixed to its counterpart equivalent parameter with $(2, 0)$.
	\item[b] Weighted unitless value for the entire fit.
	\end{tablenotes}
	\end{threeparttable}}
\end{table*}

Selected spectroscopic parameters for the two fits from this work and for the analyses of \ce{CH3C(O)CH3}, \ce{CH3 ^{13}C(O)CH3}, and \ce{^{13}CH3C(O)CH3} from Ordu et al.~\cite{Ordu2019} are shown in \autoref{tab: ParameterComparison}.
For these parameters, the two fits from this work show good agreement as the relative deviations, except for some energy tunneling parameters $\epsilon$, are below \SI{1}{\percent}. For the energy tunneling parameters the two highest relative deviations are seen for $\epsilon_{1-1}$ ($\sim\SI{13}{\percent}$) and $\epsilon_{11}$ ($\sim\SI{6}{\percent}$).
The \ce{^{13}CH3C(O)CH3} fit from Ordu et al.\ agrees for low parameters but shows already clear deviations for $\Delta_{JK}$ ($\sim\SI{5}{\percent}$) and $\delta_K$ ($\sim\SI{35}{\percent}$). Additionally, the energy tunneling parameters $\epsilon_{20}$ and $\epsilon_{02}$, which are fixed to each other in the analysis from Ordu et al., show strong deviations of about \SI{25}{\percent} each.
This is expected, as Ordu et al.\ assigned only 72 lines, suffered from misassignments, and, in contrast to this work, were severely limited in frequency and quantum number coverage.


\section{Conclusion}
\label{sec:Discussion}

The microwave spectrum of an enriched sample of acetone-\ce{^{13}C1} was recorded for the first time.
More than \SI{1}{THz} of high-resolution spectra were recorded up to \SI{1102}{\giga\hertz}.
Previous analyses of the rotational ground state were extended, with the number of assigned transitions increasing by more than a factor of 100.
The quantum number coverage was increased to $J''_\text{max}=98$ and $K''_{a,\text{max}}=44$.
Two fits are presented.
The fit up to \SI{500}{GHz} should be used for structure determination as the fit up to \SI{1100}{GHz} is expected to be effective.
The latter fit allows for astronomical searches up to the THz region, especially due to the high number of assigned lines, but caution is required for the here excluded quantum numbers above \SI{500}{GHz}.

For the analysis, the new LLWP software was used successfully.
It raised the efficiency and confidence of assignments in the dense and complicated spectrum. Additionally, LLWP facilitated the analysis by providing a multitude of important metrics.
Due to its general and unopinionated approach, molecular fingerprints of various molecules can be analyzed in various frequency ranges.

LLWP is expected to speed up and facilitate the assignment of many more complex molecular fingerprints in the future.

\section*{Acknowledgements}
This work has been supported via Collaborative Research Centre 956, sub-project B3, funded by the Deutsche Forschungsgemeinschaft, Germany (DFG; project ID 184018867) and DFG SCHL 341/15-1 (‘‘Cologne Center for Terahertz Spectroscopy’’). J.-C.G. thanks the Centre National d’Etudes Spatiales (CNES) and the ‘‘Programme National Physique et Chimie du Milieu Interstellaire‘‘ (PCMI) of CNRS/INSU with INC/INP co-funded by CEA and CNES for a grant.

\appendix
\setcounter{figure}{0}
\setcounter{table}{0}
\section{Complementary Material}
\label{sec:Appendix}

\begin{figure*}[htbp]
    \centering
    \includegraphics[width=1\textwidth]{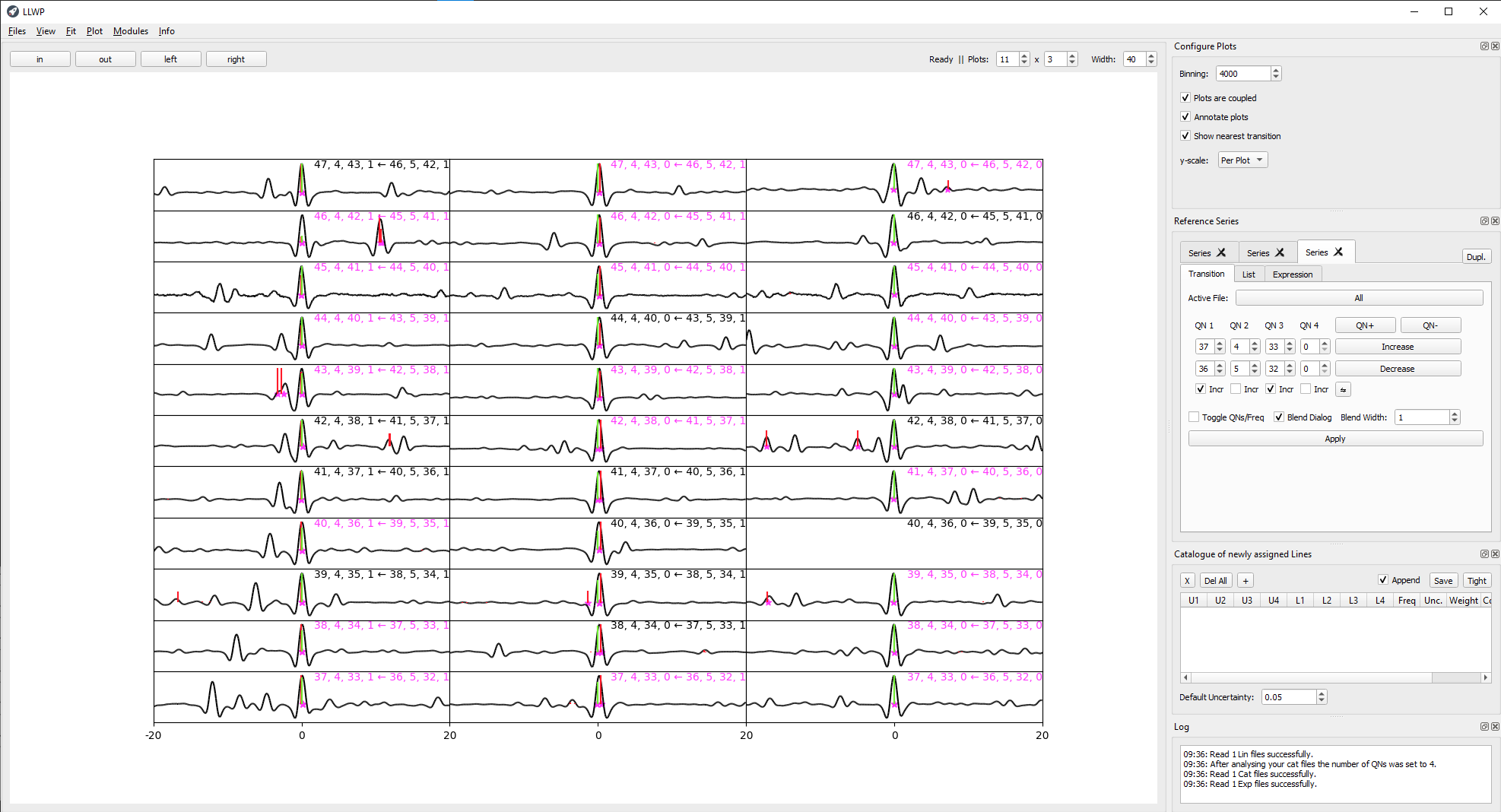}
    \caption{
		The GUI of LLWP with multiple columns.
		Each component of the triplet from \autoref{fig:GUI LLWP} is shown here in its own column.
		Other typical use cases for multiple columns include simultaneously observing different branches of infrared data or comparisons of different predictions.
	}
    \label{fig:GUI LLWP COLS}
\end{figure*}

\begin{figure*}[htbp]
     \centering
     \subfloat[Fit up to \SI{500}{GHz}]{\includegraphics[width=0.5\textwidth]{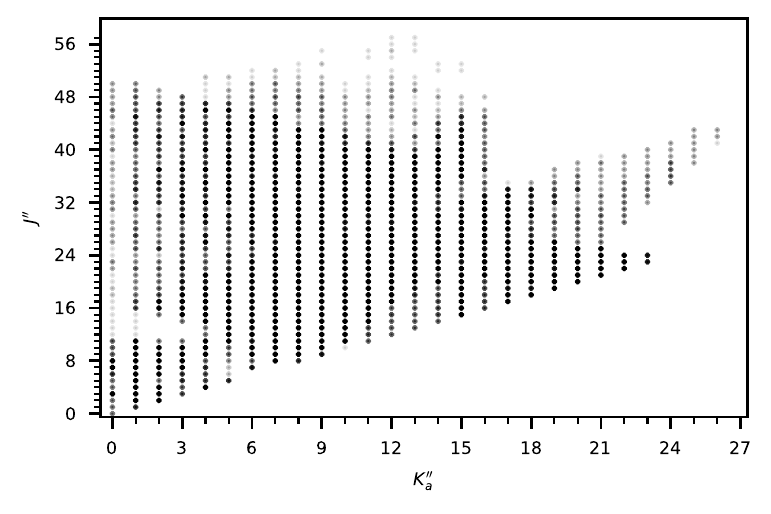}\label{fig: qncoverage500GHz}}
     \subfloat[Fit up to \SI{1100}{GHz}]{\includegraphics[width=0.5\textwidth]{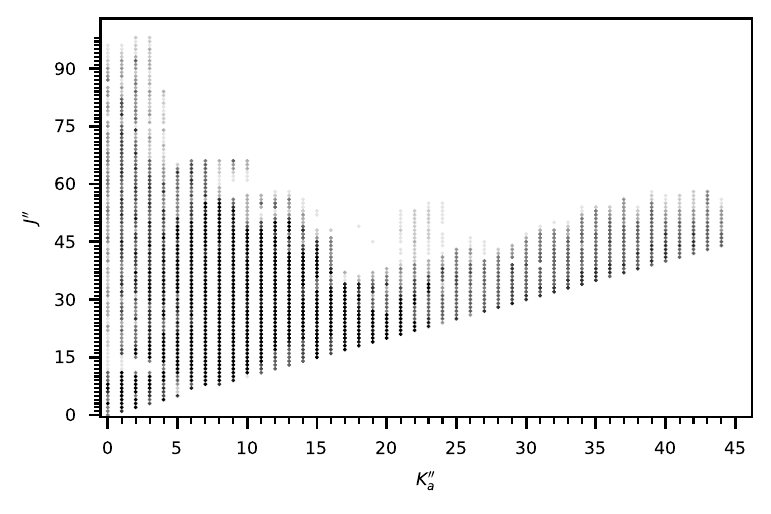}\label{fig: qncoverage1200GHz}}
     \caption{
		Quantum number coverage of the two final fits (\autoref{tab: Parameters500GHz} and \autoref{tab: Parameters1200GHz}).
		The data is plotted with high transparency, meaning the shade of each point represents the number of assigned transitions (solid black corresponds to $\geq 10$ transitions).
	}
     \label{fig: qncoverage}
\end{figure*}

\begin{figure*}[htbp]
    \centering
    \subfloat[Fit up to \SI{500}{GHz}]{\includegraphics[width=0.5\textwidth]{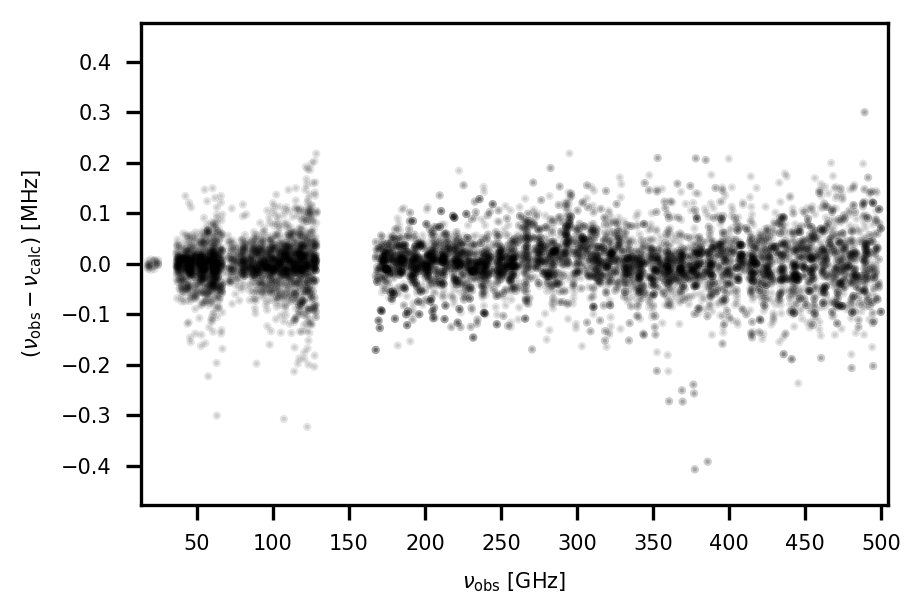}\label{fig: residuals500GHz}}
    \subfloat[Fit up to \SI{1100}{GHz}]{\includegraphics[width=0.5\textwidth]{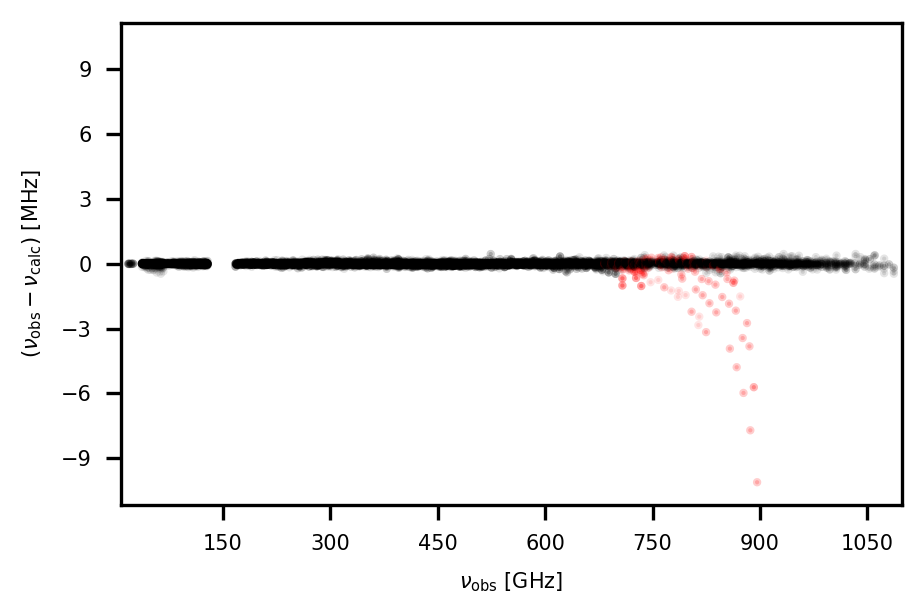}\label{fig: residuals1200GHz}}
    \caption{
		Residuals of the two final fits (\autoref{tab: Parameters500GHz} and \autoref{tab: Parameters1200GHz}).
		The assignments that were excluded from the fit up to \SI{1100}{GHz} (being the transitions with  $94 > J' > 67$ and $K_a' > 3$) are highlighted in red in \autoref{fig: residuals1200GHz}.
		They show strong deviations from the predictions and could be easily identified in LLWP's residuals window.
	}
     \label{fig: residuals}
\end{figure*}

\begin{figure*}[htbp]
    \centering
    \subfloat[Fit up to \SI{500}{GHz}]{\includegraphics[width=0.5\textwidth]{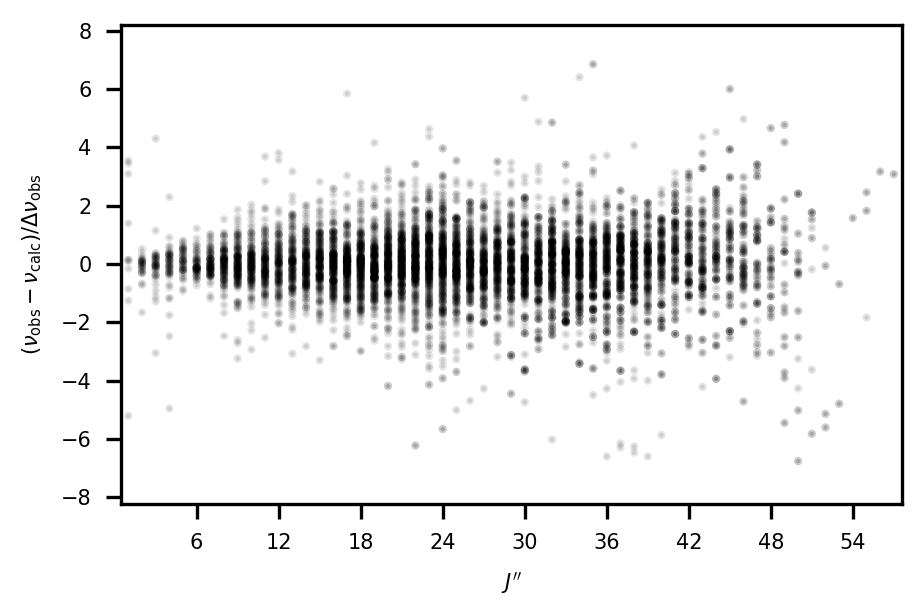}\label{fig: weightedresiduals500GHz}}
    \subfloat[Fit up to \SI{1100}{GHz}]{\includegraphics[width=0.5\textwidth]{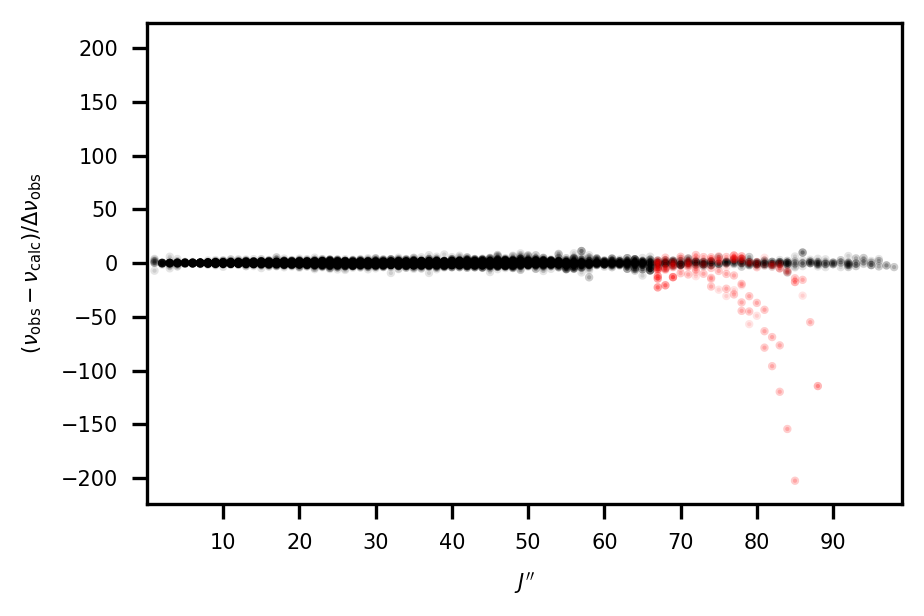}\label{fig: weightedresiduals1200GHz}}
    \caption{
		Weighted residuals of the two final fits (\autoref{tab: Parameters500GHz} and \autoref{tab: Parameters1200GHz}) against $J''$.
		The assignments that were excluded from the fit up to \SI{1100}{GHz} (being the transitions with  $94 > J' > 67$ and $K_a' > 3$) are highlighted in red in \autoref{fig: residuals1200GHz}.
	}
     \label{fig: weightedresiduals}
\end{figure*}

First, the GUI of LLWP is shown with multiple columns (\autoref{fig:GUI LLWP COLS}) to highlight the possibilities that this offers. Next, different important metrics of the final analyses are provided. The quantum number coverage plots (\autoref{fig: qncoverage}) of the two final fits give an overview of the quantum numbers of the included transitions. The residuals (\autoref{fig: residuals}) and weighted residuals (\autoref{fig: weightedresiduals}) are a measure for the goodness of the fit and highlight the need to exclude specific transitions with high deviations for the fit up to \SI{1100}{GHz}.
Additionally, the assignments per transition type are displayed in \autoref{tab:transitiontypecoverage}.

\begin{table*}[htbp]
    \centering
    \caption{
		Assignments per transition type for the two final fits (\autoref{tab: Parameters500GHz} and \autoref{tab: Parameters1200GHz}).
		Only transitions with a nonzero weight are listed here (252 transitions with zero weight together with the 15956 here listed transitions result in the 16208 assigned lines).
		For the fits, the $a$-type transitions were excluded as they are a product of label switching \autoref{sec:Acetone Fit}.
		An additional 225 lines were excluded for the fit up to \SI{1100}{\giga\hertz} (transitions with $94 > J' > 67$ and $K_a' > 3$).
	}
    \label{tab:transitiontypecoverage}
    \begin{threeparttable}    
    \begin{tabular}{rrrr|rrrr}
        \toprule
        \multicolumn{4}{c|}{Fit up to \SI{500}{\giga\hertz} (\autoref{tab: Parameters500GHz})} & \multicolumn{4}{c}{Fit up to \SI{1100}{\giga\hertz} (\autoref{tab: Parameters1200GHz})} \\
        \midrule
        $J'-J''$ & $K_a'-K_a''$ & $K_c'-K_c''$ & No. Transitions & $J'-J''$ & $K_a'-K_a''$ & $K_c'-K_c''$ & No. Transitions\\
        \midrule
        0 &  0 & -1 &    92\tnote{a} & 0 &  0 & -1 &    92\tnote{a} \\
        0 &  1 & -2 &    26          & 0 &  1 & -2 &    26 \\
        0 &  1 & -1 &  3280          & 0 &  1 & -1 &  3280 \\
        0 &  1 &  0 &    31          & 0 &  1 &  0 &    31 \\
        0 &  2 & -1 &    89\tnote{a} & 0 &  2 & -1 &    89\tnote{a} \\
        0 &  3 & -3 &     -          & 0 &  3 & -3 &     2 \\
        1 & -1 &  1 &  1868          & 1 & -1 &  1 &  2926 \\
        1 & -1 &  3 &     1          & 1 & -1 &  3 &     1 \\
        1 &  0 &  1 &  1073\tnote{a}& 1 &  0 &  1 &  3147\tnote{a} \\
        1 &  1 & -1 &   211          & 1 &  1 & -1 &   554 \\
        1 &  1 &  0 &   612          & 1 &  1 &  0 &  2023 \\
        1 &  1 &  1 &  2095          & 1 &  1 &  1 &  3783 \\
        1 &  3 & -2 &     -          & 1 &  3 & -2 &     1 \\
        1 &  3 & -1 &     1          & 1 &  3 & -1 &     1 \\
        \midrule
        \multicolumn{3}{l}{$a$-type transitions}          &  1254\tnote{a} & \multicolumn{3}{l}{$a$-type transitions} &  3328\tnote{a} \\
        \multicolumn{3}{l}{$b$-type transitions}          &  7456          & \multicolumn{3}{l}{$b$-type transitions} & 10547          \\
        \multicolumn{3}{l}{$c$-type transitions}          &   669          & \multicolumn{3}{l}{$c$-type transitions} &  2081          \\
        \midrule
        \multicolumn{3}{l}{Total transitions}             &  9379          & \multicolumn{3}{l}{Total transitions}    & 15956          \\
        \bottomrule
        \end{tabular}
        \begin{tablenotes}\footnotesize
        \item \textbf{{Note.}} Primes indicate the upper level and double primes indicate the lower level.
        \item[a] $a$-type transitions were excluded from all fits, see \autoref{sec:Acetone Fit}
        \end{tablenotes}
    \end{threeparttable}
\end{table*}

\biboptions{comma,sort&compress}
\bibliographystyle{elsarticle-num}
\bibliography{resources/bibliography}

\end{document}